\documentclass[12pt]{iopart}

\usepackage{iopams}

\usepackage[dvips]{color}
\usepackage{graphicx}
\usepackage{subfigure}
\usepackage{epsfig}
\usepackage{float}

\def\iotabar{\lower3pt\hbox{$\mathchar'26$}\mkern-7mu\iota}
\newcommand {\aplt} {\ {\raise-.5ex\hbox{$\buildrel<\over\sim$}}\ }
\newcommand{\dd}{\mbox{d}}

\begin{document}

\title[Fluid limit of Continuous Time Random Walks on the
circle]{Continuous Time Random Walks in periodic systems: fluid limit
and fractional differential equations on the circle}

\author{I Calvo$^1$\footnote{Corresponding author. E-mail: {\tt
ivan.calvo@ciemat.es}}, B A Carreras$^2$, R S\'anchez$^3$ and B Ph van
Milligen$^1$}

\address{$^1$ Laboratorio Nacional de Fusi\'on, Asociaci\'on
EURATOM-CIEMAT, E-28040 Madrid, Spain}

\address{$^2$ BACV Solutions Inc., Oak Ridge, TN 37830,
U.S.A.}

\address{$^3$ Fusion Energy Division, Oak Ridge National Laboratory,
Oak Ridge, TN 37831, U.S.A.}


\begin{abstract}
In this article, the continuous time random walk on the circle is
studied. We derive the corresponding generalized master equation and
discuss the effects of topology, especially important when L\'evy
flights are allowed. Then, we work out the fluid limit equation,
formulated in terms of the periodic version of the fractional
Riemann-Liouville operators, for which we provide explicit
expressions. Finally, we compute the propagator in some simple
cases. The analysis presented herein should be relevant when
investigating anomalous transport phenomena in systems with periodic
dimensions.
\end{abstract}


\section{Introduction}

Continuous Time Random Walks (CTRWs) (\cite{MonWei}, \cite{SchLax})
are models describing the motion of individual particles (or any other
conserved quantity) probabilistically.  Since their introduction they
have found many applications in physics, among other reasons, due to
the fact that they allow exploring possibilities which go beyond the
classical paradigm of diffusive transport. Their simplest realization,
known as a separable CTRW, is defined in terms of two probability
distribution functions (pdfs): the step-size pdf, $p$, and the
waiting-time pdf, $\psi$. It is well-known that when the step-size pdf
is Gaussian and the waiting-time pdf is exponential, the fluid limit
(in which only long-time, large-distance information is retained)
yields diffusive equations. These choices intrinsically imply the
existence of finite characteristic length and time scales associated
to the underlying microscopic transport mechanism.

However, there is a large number of physical, biological and social
systems ~\cite{HauKeh,BouGeo,Metzler00,Zaslavsky02} in which the
dominant transport mechanism lacks either a well-defined
characteristic length scale or a well-defined characteristic timescale
or both.  Instead, the characteristic transport scales diverge with
the system size and/or lifespan. For instance, this happens whenever
transport takes place via avalanches. Transport events as a result
have a maximum size that is only limited by the system size $L$, and a
characteristic size that diverges with some power of $L$.  This
situation has been encountered, to list just a few examples, while
investigating the transport of both particles and energy out of
magnetically-confined fusion
plasmas~\cite{Newman96,Carreras96,Sanchez01}, in the propagation of
forest fires~\cite{Drossel92}, earthquakes~\cite{Shaw92}, solar
flares~\cite{Lu91} and the transport of magnetic vortices in type-II
superconductors~\cite{Field95}. It is possible to construct CTRW
models that lack characteristic scales by choosing $p$ and $\psi$ from
the family of stable L\'evy distributions~\cite{Taqqu}. The fluid
limit is then expressed in terms of transport equations that contain
fractional differential operators~\cite{Podlubny}, which are
essentially non-local and non-Markovian. There is evidence, both
numerical and experimental, that models generated in this manner
provide a reasonably effective description of transport for some of
the aforementioned
problems~\cite{Metzler00,Zaslavsky02,Carreras01,Castillo04}.

Although the formulation of CTRW models in finite systems is
straightforward~\cite{MilSanCar04,Sanchez05}, their fluid limit
equations and associated propagators are much harder to obtain
than in infinite domains. In this work we will study the
formulation of CTRWs on a finite one-dimensional system with
periodic boundary conditions (the unit circle) and work out its
fluid limit in a rigorous way for an arbitrary stable L\'evy
distribution.

We envision that the formalism introduced in this paper might find
application in systems in which particles or energy are amenable to
trapping and/or avalanching processes along a periodic direction. A
possible example might be the investigation of heat transport in
single-wall carbon nanotube (SWNT) nanorings, which have gained recent
attention due to their interesting transport
properties~\cite{Kasai03}, such as Aharonov-Bohm effects,
magnetotransport or establishment of persistent currents. Some
molecular dynamical simulations have found that linear SWNTs seem to
exhibit heat conductivities divergent with the nanotube length, which
suggests that a characteristic length scale might be
lacking~\cite{Maruyama02}. If these observations apply to nanorings as
well, the periodic CTRW here presented might provide an adequate model
for their effective description. Other possible applications might be
to the modeling of some types of transport in turbulent plasmas
confined in a (multi-)periodic system, such as those found in the
(quasi-spherical) sun or in Earth-based (toroidal) magnetic
confinement devices with interest for fusion energy
production~\cite{Caletal07}.

The paper is organized as follows. In Section \ref{sec:CTRWperGME} we
derive the generalized master equation (GME) for a time-translational
invariant, separable CTRW on the circle. In Section
\ref{sec:FluidLimit} we work out the fluid limit of such family of
CTRWs introducing, in particular, the appropriate form of the
Riemann-Liouville operators on the circle. The computation of the
propagator of the fluid limit equations in the homogeneous case is
presented in Section \ref{sec:propagator}. We discuss its asymptotic
behaviour and give analytical solutions for $\alpha=1,2$ in the
Markovian case. Section \ref{sec:conclusions} is devoted to the
conclusions. Finally, \ref{sec:appLevy} and \ref{sec:FDOrealine} give
a brief survey on stable L\'evy distributions and fractional calculus,
respectively.

\section{CTRW on the circle. Generalized Master Equation}
\label{sec:CTRWperGME}

We consider that the dynamics takes place on the unit circle, $S^1$,
parameterized by $\theta\in[0,2\pi)$. Sometimes it is useful to view
the unit circle as $S^1\cong{\mathbb R}/2\pi$, where the quotient is
taken with respect to the equivalence relation $x\sim x+2\pi$ for any
$x\in\mathbb R$. Functions on $S^1$ can then be identified with the
set of periodic functions on $\mathbb R$ with period
$2\pi$. Explicitly, given $f:S^1\rightarrow \mathbb R$, its extension
to a periodic function on $\mathbb R$ is
\begin{equation}
\begin{array}{cccc}
\breve f:&\mathbb R&\rightarrow &\mathbb R\\[8pt]
{}& x & \mapsto& f([x]),
\end{array}
\end{equation}
where $[x]$ is the (unique) representative of the equivalence class of
$x$ which belongs to $[0,2\pi)$. Obviously, if $x\in[0,2\pi)$, $\breve
f(x)=f(x)$. Finally, every function $f\in S^1$ considered in this
paper is assumed to have a Fourier series expansion,
\begin{equation}
f(\theta)= \frac{1}{2\pi}\sum_{m=-\infty}^\infty f_m
e^{-im\theta},
\end{equation}
with coefficients given by the formula:
\begin{equation}
f_m=\int_0^{2\pi} f(\theta)e^{im\theta}\dd\theta.
\end{equation}

As mentioned above, we restrict ourselves to the case of separable
CTRWs with time-translational invariance. This means that the CTRW is
defined by two pdfs, $p(\Delta,\theta)$ and $\psi(\tau,\theta)$,
called the {\it step-size} pdf and the {\it waiting-time} pdf,
respectively. Here, $\Delta\in\mathbb R$, $\tau\geq 0$ and $\theta\in
[0,2\pi)$. $p$ gives the probability that a particle located at
$\theta$ jumps to $\theta+\Delta$ after having waited at $\theta$ for
a lapse of time $\tau$. Since $\Delta$ is unbounded, this may imply an
arbitrary number of turns around the circle. Finally, $p$ and $\psi$
verify the normalization conditions,
\begin{eqnarray}
&&\int_{-\infty}^{\infty}p(\Delta,\theta)\dd\Delta=1,\
\forall\theta\in[0,2\pi),\\[8pt]
&&\int_0^\infty\psi(\tau,\theta)\dd\tau =1, \ \forall\theta\in
[0,2\pi).
\end{eqnarray}

We denote by $n(\theta,t)$ the density of particles normalized to the
total number of particles, i.e. $\int_0^{2\pi}n(\theta,t)\dd\theta=1,
\ \forall t$. The GME for the time evolution of $n(\theta,t)$
corresponding to this class of CTRWs can be derived as
follows~\cite{KreMonSch,MilSanCar04}. First, note that the normalized
density of particles is equal to the probability of finding one
particle at $\theta$ at time $t$. This probability can be rewritten
as:
\begin{equation}
n(\theta, t) = \int_0^t Q(\theta, t') \eta(\theta, t-t')\dd t',
\end{equation}
where $Q(\theta, t')$ gives the probability that the walker arrives at
$\theta$ at time $t'<t$, and $\eta(\theta, \tau)$ gives the
probability of staying at $\theta$ for a lapse of time
$\tau$. Clearly, one has:
\begin{equation}
\eta(\theta,t)=1-\int_0^t\psi(\theta,\tau)\dd\tau.
\end{equation}
Then, we rewrite:
\begin{equation}
Q(\theta,t) = \sum_{j=0}^\infty Q^j(\theta, t),
\end{equation}
where $Q^j(\theta,t)$ denotes the probability that a single walker
arrives at $\theta$ at time $t$ by performing exactly $j$
jumps. Obviously, for an arbitrary CTRW the function $Q^j(\theta,t)$
satisfies the following recurrence relation:
\begin{equation}\label{eq:Qj}
\hspace{-1cm}Q^j(\theta,t)=\sum_{m=-\infty}^\infty\int_0^{2\pi}\dd\theta'
\int_0^t\dd t'p(\theta-\theta'+2\pi m,\theta')
\psi(t-t',\theta')Q^{j-1}(\theta',t'),
\end{equation}
where the sum in front of the integral explicitly accounts for the
fact that particles can arrive at $\theta$ from $\theta'$ through
jumps of length $|\theta-\theta'+2\pi m|$, $m\in{\mathbb Z}$.
$Q^0(\theta, t) = \delta(\theta)\delta(t)$ provides the needed initial
condition. Adding over all possible values of $j$, we find an equation
for $Q(\theta, t)$:
\begin{equation}\label{eq:Qj1}
\hspace{-2cm}Q(\theta,t) -\delta(\theta)\delta(t)
=\sum_{m=-\infty}^\infty\int_0^{2\pi}\dd\theta' \int_0^t\dd
t'p(\theta-\theta'+2\pi m,\theta')
\psi(t-t',\theta')Q(\theta',t').
\end{equation}

We proceed now to carry out the previously described periodic
extension to the real line. Trivially, $\breve p$, $\breve \psi$ and
$\breve Q$ also satisfy Eq.~(\ref{eq:Qj1}). In $\mathbb R$, we can make
the change of variables ${\tilde\theta}=\theta'-2\pi m$ and use the
periodicity properties of $\breve p$, $\breve\psi$ and
$\breve Q$, together with the additivity of the integral with respect
to the interval of integration, and convert Eq.~(\ref{eq:Qj1}) into:
\begin{equation}\label{eq:Qj2}
\hspace{-1cm}\breve Q(\theta,t) -\delta(\theta)\delta(t) =
\int_{-\infty}^{\infty}\dd\theta' \int_0^t\dd t'\breve
p(\theta-\theta',\theta') \breve\psi(t-t',\theta')\breve
Q(\theta',t').
\end{equation}
This equation is formally identical to Eq.~(12) from Ref.
\cite{MilSanCar04}, which appears in the derivation of the GME in an
infinite system. The main difference is that $\breve Q(\theta, t)$,
whilst positive, is no longer a probability density in $\mathbb R$,
since as a periodic function it is no longer integrable. This does not
invalidate the procedure, because the convolution theorem still
applies. Consequently, we can take the final result: Eq.~(24) of
Ref. \cite{MilSanCar04}. Namely,
\begin{equation}\label{eq:GMEinf}
\fl \partial_t \breve n(\theta,t)= \int_0^t\dd t'\int_{-\infty}^\infty\dd
\theta'\breve p(\theta-\theta',\theta')\phi(\theta',t-t')\breve n(\theta',t')-
\int_0^t\phi(\theta,t-t')\breve n(\theta,t')\dd t',
\end{equation}
where $\phi$ is also a periodic function defined by:
\begin{eqnarray}\label{eq:defetaphi}
\phi(\theta,t)={\cal L}^{-1}\left[\frac{{\cal
L}[\breve\psi(\theta,t)]}{{\cal L}[\breve\eta(\theta,t)]}\right],
\end{eqnarray}
and ${\cal L}[\cdot]$ stands for the Laplace transform with respect to
$t$.

\begin{figure}[htp]
\begin{center}
\subfigure[Gaussian]{
\label{fig:compare_gauss}
\resizebox{2.7in}{!}{\includegraphics[angle=-90]{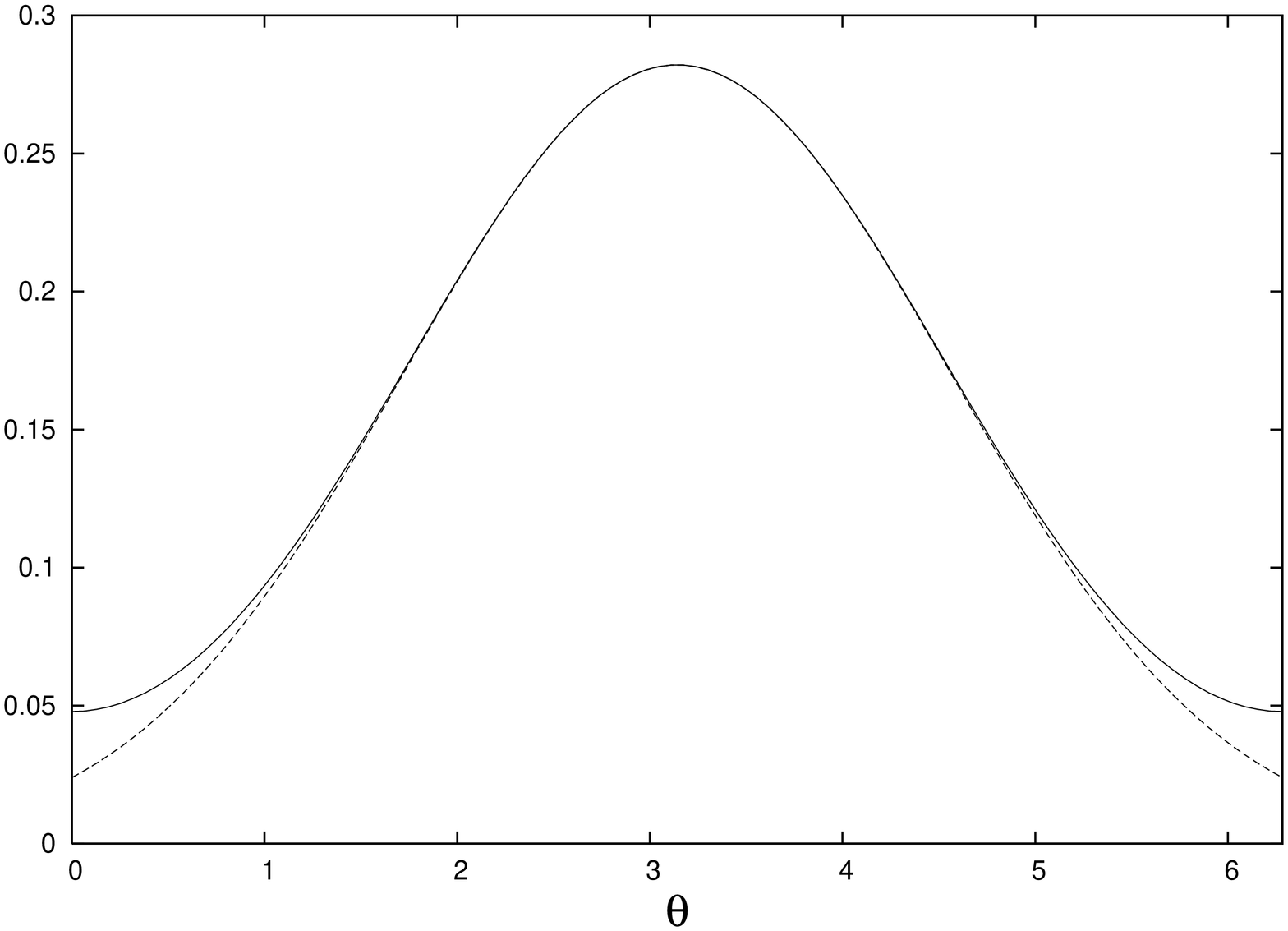}}
}
\subfigure[Cauchy]{
\label{fig:compare_cauchy}
\resizebox{2.7in}{!}{\includegraphics[angle=-90]{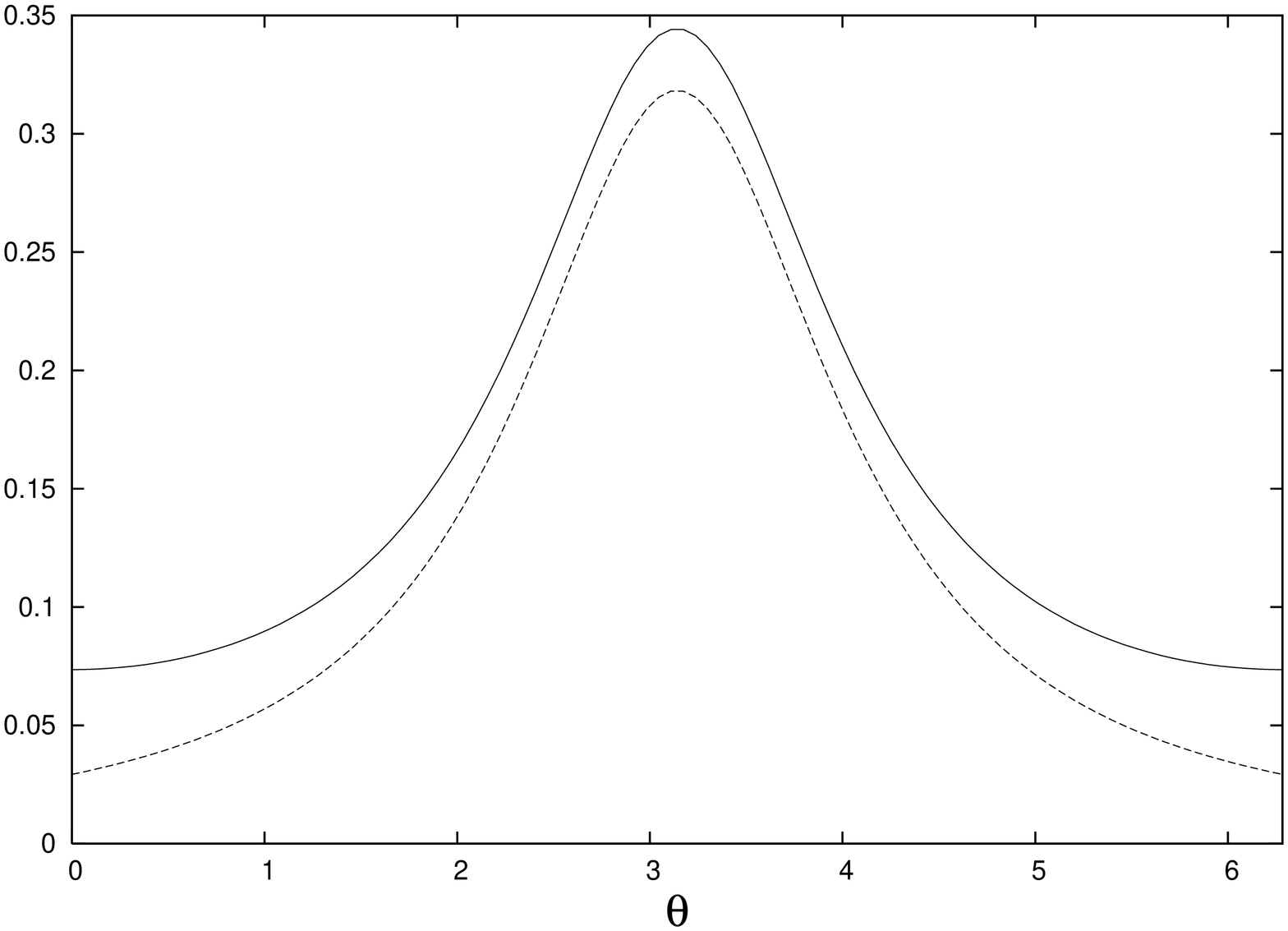}}
}
\caption{Comparison of $p$ (dashed) and its associated $\bar p$
(solid). (a) $p$ is a symmetric Gaussian distribution
(i.e. $\alpha=2$). (b) $p$ is a symmetric Cauchy distribution
(i.e. $\alpha=1$). In both subfigures, $p$ has $\sigma=1$ and is
centered at $\theta=\pi$.}
\label{fig:compare_pwithpbar}
\end{center}
\end{figure}

Again, $\breve n(\theta,t)$ is not a probability density function in
$\mathbb R$ even when positive because it is periodic and thus
non-integrable. We need to restrict the integrals to $[0,2\pi)$ to
recover the probabilistic interpretation. To do so, we simply reverse
the manipulations which led us from (\ref{eq:Qj1}) to (\ref{eq:Qj2}),
recasting the GME (\ref{eq:GMEinf}) as
\begin{equation}\label{eq:GMEper}
\fl \partial_t n(\theta,t)= \int_0^t\dd t'\int_{0}^{2\pi}\dd
\theta'{\bar p}(\theta-\theta',\theta')\phi(\theta',t-t')n(\theta',t')-
\int_0^t\phi(\theta,t-t')n(\theta,t')\dd t',
\end{equation}
with
\begin{equation}\label{eq:barp}
\bar p(\theta,\theta') = \sum_{m=-\infty}^\infty
p(\theta+2\pi m,\theta').
\end{equation}
From this expression it is obvious that $\bar p$ is positive-definite
and periodic in its first argument, $\bar p(0,\theta')=\bar
p(2\pi,\theta')$, for any $\theta'$. Furthermore, the normalization of
$p$ in $\mathbb R$ implies the normalization of $\bar p$ in $S^1$,
\begin{equation}
\int_0^{2\pi}\bar p(\theta,\theta')\dd\theta =1, \ \forall\theta'\in[0,2\pi),
\end{equation}
so that $\bar p$ is a probability distribution on the circle. Hence,
the GME (\ref{eq:GMEper}) is written in terms of functions on $S^1$
and makes manifest that the dynamics of $n(\theta,t)$ is well-defined
on the circle. It is worth pointing out that (\ref{eq:barp}) expresses
the fact that particles can wind around the circle an arbitrary number
of times, and hence is a consequence of the non-trivial topology of
the manifold on which the CTRW takes place. Actually, $\bar p$ shows
up as the effective step-size pdf on the circle and is related to the
original step-size pdf, $p$, through (\ref{eq:barp}). This relation is
in fact the \emph{ballooning transform} that was introduced in the
plasma physics literature in the late 70s to deal with PDEs formulated
on toroidal geometries~\cite{Connor}. Figs. \ref{fig:compare_gauss}
and \ref{fig:compare_cauchy} compare $p$ and $\bar p$ when $p$ is a
Gaussian and a Cauchy distribution (see \ref{sec:appLevy}),
respectively. As expected, the effect of topology is much more
relevant for L\'evy distributions with algebraic tails than for the
Gaussian.

\section{Fluid limit of the GME. Fractional differential operators on
the circle}\label{sec:FluidLimit}

Since we will be especially interested in the fluid limit of the CTRW
when $p$ and/or $\psi$ are L\'evy distributions (which only have a
finite number of finite integer moments), we cannot carry out the
calculation by the standard method of expanding the GME using moments
of these distributions. The appropriate procedure involves going to
the Fourier (for space) and Laplace (for time) domains.  The fluid
limit, which implies retaining only the dynamical information
pertaining to the longer times and larger distances, corresponds to
taking the limits $k\rightarrow 0, s\rightarrow 0$, where $k$ and $s$
stand respectively for the Fourier and Laplace variables.

Since $\bar p$ is periodic with period $2\pi$, we can expand it in a
Fourier series with respect to its first argument:
\begin{equation}
\bar p(\theta,\theta')= \frac{1}{2\pi}\sum_{m=-\infty}^\infty \bar
p_m(\theta') e^{-im\theta}.
\end{equation}
 Recalling the definition of $\bar p$ in terms of $p$, the
coefficients of this expansion are:
\begin{eqnarray}
\fl {\bar p}_m(\theta')&\hspace{-1cm}=&\int_{0}^{2\pi}\bar
p(\theta,\theta')e^{im\theta}\ \dd\theta =
\sum_{m'=-\infty}^\infty\int_{0}^{2\pi} p(\theta+2\pi
m',\theta')e^{im\theta}\ \dd \theta\cr &\hspace{-1cm}=&
\sum_{m'=-\infty}^\infty\int_{2\pi m'}^{2\pi(m'+1)}
p(\tilde\theta,\theta')e^{im\tilde\theta}e^{-i2\pi mm'}\dd
\tilde\theta = \sum_{m'=-\infty}^\infty\int_{2\pi
m'}^{2\pi(m'+1)}p(\tilde\theta,\theta')e^{im\tilde\theta}\dd\tilde\theta\cr
&\hspace{-1cm}=& \int_{-\infty}^\infty
p(\tilde\theta,\theta')e^{im\tilde\theta}\dd\tilde\theta=\hat
p(m;\theta').
\end{eqnarray}
where in the third equality we have performed the change of
variables $\tilde\theta=\theta+2\pi m'$ and in the fourth equality
we have used that $e^{-i2\pi mm'}=1$. Hence, the $m$-th
coefficient in the Fourier expansion of $\bar p$ with respect to
its first argument is exactly the value of the Fourier transform
of $p$ with respect to its first argument at $k=m$. This relation
is characteristic of the aforementioned ballooning
transform~\cite{Connor}.

Following Ref.~\cite{MilCarSan05}, we proceed by inserting
this expansion in (\ref{eq:GMEper}),
\begin{eqnarray}
\hspace{-1cm}\partial_t
n(\theta,t)&=&\frac{1}{2\pi}\sum_{m'=-\infty}^\infty \int_0^t\dd
t'\int_{0}^{2\pi}\dd \theta'\phi(\theta',t-t')\bar
p_{m'}(\theta') e^{-im'(\theta-\theta')}n(\theta',t')\cr &-&
\int_0^t\phi(\theta,t-t')n(\theta,t')\dd t'.
\end{eqnarray}
Now, we expand both sides in Fourier series with respect to $\theta$
and equate coefficients,
\begin{eqnarray}
\fl\partial_t
n_m(t)&\hspace{-1cm}=&\frac{1}{2\pi}\sum_{m'=-\infty}^\infty
\int_0^{2\pi}\dd\theta e^{i(m-m')\theta} \int_0^t\dd t'\int_{0}^{2\pi}\dd
\theta'\phi(\theta',t-t')\bar p_{m'}(\theta')
e^{im'\theta'}n(\theta',t')\cr &\hspace{-1cm}-&
\int_0^{2\pi}\dd\theta
e^{im\theta}\int_0^t\phi(\theta,t-t')n(\theta,t')\dd t'.
\end{eqnarray}
Since $\int_0^{2\pi}e^{i(m-m')\theta}\dd\theta=2\pi\delta_{mm'}$ and $\bar
p_m(\theta')=\hat p(m;\theta')$ we obtain:
\begin{equation}\label{eq:GMEperFourierinhom}
\partial_t
n_m(t)=\int_0^t\dd t'\int_{0}^{2\pi}\dd
\theta'\phi(\theta',t-t')\left(\hat p(m;\theta')-1\right)
e^{im\theta'}n(\theta',t').
\end{equation}

Let $\Lambda(k;\theta)$ be the characteristic exponent of
$p(\Delta,\theta)$, i.e. $\hat
p(k;\theta)=\exp\Lambda(k;\theta)$. The fluid limit is defined by
taking $\hat p(k;\theta)\approx 1+\Lambda(k;\theta)$. Therefore,
Eq.~(\ref{eq:GMEperFourierinhom}) in the fluid limit is written as
\begin{equation}\label{eq:GMEperFourierinhomfl}
\partial_t n_m(t)=\int_{0}^{2\pi}\dd \theta'\Lambda(m;\theta')
e^{im\theta'}\int_0^t\dd t'\phi(\theta',t-t')n(\theta',t').
\end{equation}

At this point we need to make choices for the step-size and
waiting-time pdfs. A natural choice is to take stable L\'evy
distributions, which satisfy the central limit
theorem~\cite{Taqqu}. Thus, following Ref.~\cite{SanCarMil05}, we will
assume first that the waiting-time pdf is a positive extremal L\'evy
distribution (see Appendix A). Performing a (non-analytic, in general)
expansion of the Laplace transform of $\psi$ around $s=0$ and keeping
the lowest order terms:
\begin{equation}
\tilde\psi(s)\approx 1-A_\gamma^{-1}\tau^\gamma s^\gamma,\quad
\gamma\in(0,1],
\end{equation}
where $A_\gamma=\cos(\pi\gamma/2)$ if $\gamma\in(0,1)$ and
$A_\gamma=1$ if $\gamma=1$, which corresponds to an exponential
waiting-time pdf. The convolution in the second integral of
(\ref{eq:GMEperFourierinhomfl}) in the fluid limit yields:
\begin{equation}
{\cal L}\left[\int_0^t\dd t'\phi(\theta',t-t')n(\theta',t')\right]\approx
A_\gamma\tau^{-\gamma}s^{1-\gamma}\tilde n(\theta',s)
\end{equation}
where we have made use of (\ref{eq:LevyextremalLaplace}). Then,
Eq.~(\ref{eq:GMEperFourierinhomfl}) in Laplace space reads:
\begin{equation}\label{eq:GMEperFourierinhomflLaplace}
s\tilde n_m(s)-n_m(0)=
A_\gamma\tau^{-\gamma}s^{1-\gamma}\int_{0}^{2\pi}
\Lambda(m;\theta')\tilde n(\theta';s) e^{im\theta'}\dd \theta'.
\end{equation}
Now, multiplying both sides by $s^{\gamma-1}$, using (\ref{eq:Caputo})
and (\ref{eq:CaputoLaplace}) and Laplace inverting:
\begin{equation}\label{eq:GMEperFourierinhomfl1}
\frac{\dd^{\gamma}}{\dd t^{\gamma}} n_m(t)=
A_\gamma\tau^{-\gamma}\int_{0}^{2\pi}
\Lambda(m;\theta')n(\theta',t)e^{im\theta'}\dd \theta'.
\end{equation}

As for the spatial part, we will consider the case in which
$\Lambda(k;\theta)$ is the characteristic exponent of a general L\'evy
distribution (see \ref{sec:appLevy}) with parameters dependent on
$\theta$:
\begin{equation}\label{eq:defLevyvar}
\fl \Lambda(k;\theta)=
\left\{
\begin{array}{cc}
-\sigma^\alpha(\theta)|k|^\alpha
\left[1-i\beta(\theta)\mbox{sign}(k)\tan(\frac{\pi\alpha}{2})\right]+i\mu(\theta)k
&\hspace{0.4cm}\alpha\in(0,1)\cup(1,2],\\[8pt]
\hspace{-5.2cm}-\sigma(\theta)|k|+i\mu(\theta) k &
\hspace{-1.6cm}\alpha = 1,
\end{array}
\right.
\end{equation}
restricting ourselves to $\beta\equiv 0$ for $\alpha=1$.

Let us work out the case $\alpha\neq
1$. Eq.~(\ref{eq:GMEperFourierinhomfl1}) becomes
\begin{eqnarray}\label{eq:GMEperFourierinhomflalphaneq1}
\fl\frac{\dd^{\gamma}}{\dd t^{\gamma}}
n_m(t)&=&-A_\gamma\tau^{-\gamma}\Bigg(\int_{0}^{2\pi}
\sigma^\alpha(\theta')|m|^\alpha
\left[1-i\beta(\theta')\mbox{sign}(m)\tan\left(\frac{\pi\alpha}{2}\right)\right]e^{im\theta'}n(\theta',t)\dd \theta'\cr
\fl&+&\int_{0}^{2\pi}im\mu(\theta')
e^{im\theta'}n(\theta',t)\dd \theta'\Bigg).
\end{eqnarray}
or equivalently,
\begin{equation}\label{eq:GMEperFourierinhomflalphaneq2}
\fl\frac{\dd^{\gamma}}{\dd
t^{\gamma}}
n_m(t)=-\frac{A_\gamma}{\tau^{\gamma}}\left[|m|^\alpha\left(\sigma^\alpha
 n\right)_m-i\tan\left(\frac{\pi\alpha}{2}\right)|m|^\alpha\mbox{sign}(m)\left(\beta\sigma^\alpha n\right)_m
+ im\left(\mu n\right)_m\right](t).
\end{equation}

The inverse Fourier transform of
Eq.~(\ref{eq:GMEperFourierinhomflalphaneq2}) is the desired fluid
limit of the CTRW on the circle. The analogue of
(\ref{eq:GMEperFourierinhomflalphaneq2}) in an infinite system is
Fourier inverted by means of the Riemann-Liouville operators defined
in Appendix B. The following theorem (see \cite{Samko} for related
results) yields the suitable generalization of the Riemann-Liouville
operators on the circle.

\vskip 0.3cm

{\bf Theorem:} {\it Let $f$ be a sufficiently well-behaved function on
$S^1$, $\alpha\in(0,1)\cup(1,2)$, and define the operators}
\begin{eqnarray}\label{eq:RLpertheorem}
\fl&&{}_{0}{\cal D}^\alpha f(\theta)=
{}_{0}{D}^\alpha f(\theta) +
\frac{1}{(2\pi)^{\alpha+1}\Gamma(-\alpha)}\int_{0}^{2\pi}f(\theta')\zeta\left(1+\alpha,1+\frac{\theta-\theta'}{2\pi}\right)\dd\theta'\\[8pt]
\fl&&{}^{2\pi}{\cal D}^\alpha f(\theta)=
{}^{2\pi}{D}^\alpha f(\theta) -\frac{1}{(2\pi)^{\alpha+1}\Gamma(-\alpha)}
\int_{0}^{2\pi}f(\theta')\zeta\left(1+\alpha,1+\frac{\theta'-\theta}{2\pi}\right)\dd\theta',\nonumber
\end{eqnarray}
{\it where ${}_{0}{D}^\alpha$ and ${}^{2\pi}{D}^\alpha$ are the
Riemann-Liouville operators defined in (\ref{eq:RL}), $q$ is the
smallest integer greater than $\alpha$ and $\zeta$ is the Hurwitz zeta
function,}
\begin{equation}
\zeta(s,a)=\sum_{m=0}^\infty\frac{1}{(m+a)^s}\ , \quad \mbox{Re}(s)>1, \
m+a\neq 0.
\end{equation}
{\it Then,}
\begin{equation}
\left({}_{0}{\cal D}^\alpha f\right)_m=(-im)^\alpha f_m,\quad
\left({}^{2\pi}{\cal D}^\alpha f\right)_m=(im)^\alpha f_m.
\end{equation}

\vskip 0.3cm

{\it Proof:}

\vskip 0.3cm

First, notice that it is enough to show that ${}_{0}{\cal D}^\alpha
e^{im\theta}=(im)^\alpha e^{im\theta}$ and ${}^{2\pi}{\cal D}^\alpha
e^{im\theta}=(-im)^\alpha e^{im\theta}$. Let us give the proof for
${}_{0}{\cal D}^\alpha$. By the definition of the Hurwitz zeta
function and recalling that $\Gamma(z+1)=z\Gamma(z)$,
\begin{eqnarray}\label{eq:theorem1}
\hspace{-1.5cm}{}_{0}{\cal D}^\alpha f(\theta)& =& {}_{0}{D}^\alpha
f(\theta) + \frac{1}{\Gamma(-\alpha)}\sum_{m=0}^\infty
\int_{0}^{2\pi}\frac{f(\theta')}{(2\pi(m+1)+\theta-\theta')^{1+\alpha}}\dd\theta'\cr
&=&{}_{0}{D}^\alpha
f(\theta) + \frac{1}{\Gamma(q-\alpha)}\sum_{m=0}^\infty\frac{\dd^q}{\dd\theta^q}
\int_{0}^{2\pi}\frac{f(\theta')}{(2\pi(m+1)+\theta-\theta')^{1+\alpha-q}}\dd\theta'.
\end{eqnarray}
Performing the change of variables $\tilde\theta=\theta'-2\pi(m+1)$,
and using the periodicity of $\breve f$ we have:
\begin{eqnarray}
&&\hspace{-1.5cm}\sum_{m=0}^{\infty}\frac{\dd^q}{\dd\theta^q}
\int_0^{2\pi}
\frac{f(\theta')}{(2\pi(m+1)+\theta-\theta')^{1+\alpha-q}}\dd\theta'=\cr
&&\hspace{-1.5cm}\sum_{m=0}^{\infty}\frac{\dd^q}{\dd\theta^q}\int_{-2\pi(m+1)}^{-2\pi m}
\frac{\breve f(\tilde\theta+2\pi(m+1))}{(\theta-\tilde\theta)^{1+\alpha-q}}
\dd\tilde\theta =\lim_{a\to -\infty}
\frac{\dd^q}{\dd\theta^q}\int_{a}^{0}
\frac{\breve f(\tilde\theta)}{(\theta-\tilde\theta)^{1+\alpha-q}}\dd\tilde\theta.
\end{eqnarray}
Recalling the definition (\ref{eq:RL}) of the Riemann-Liouville
operators we immediately obtain that
\begin{equation}
{}_{0}{\cal D}^\alpha f(\theta)={}_{-\infty}{D}^\alpha \breve f(\theta).
\end{equation}
Since ${}_{-\infty}{D}^\alpha e^{im\theta}=(im)^\alpha e^{im\theta}$,
the result follows. An analogous calculation shows that
${}^{2\pi}{\cal D}^\alpha e^{im\theta}=(-im)^\alpha
e^{im\theta}$. $\,$\hfill$\Box$\break

Let us go back to Eq.~(\ref{eq:GMEperFourierinhomflalphaneq2}). Making
use of the Theorem and the identities
\begin{equation}
|m|^\alpha=\frac{(im)^\alpha+(-im)^\alpha}{2\cos(\pi\alpha/2)},
 \quad\quad
 |m|^\alpha\mbox{sgn}(m)=\frac{(im)^\alpha-(-im)^\alpha}{2i\sin(\pi\alpha/2)},
\end{equation}
for $\alpha\in(0,1)\cup(1,2)$, we can Fourier-invert
Eq.~(\ref{eq:GMEperFourierinhomflalphaneq2}) in terms of the operators
defined in (\ref{eq:RLpertheorem}):
\begin{eqnarray}\label{eq:GMEperFourierinhomflalphaneq3}
\hspace{-2cm}\frac{\dd^{\gamma}}{\dd t^{\gamma}} n(\theta,t)&=&
-\frac{A_\gamma}{2\tau^{\gamma}\cos(\pi\alpha/2)} \left[{}_{0}{\cal
D}^\alpha_\theta\left((1+\beta)\sigma^\alpha n\right)+ {}^{2\pi}{\cal
D}^\alpha_\theta\left((1-\beta)\sigma^\alpha n\right)\right](\theta,t)\cr
&+&\frac{A_\gamma}{\tau^{\gamma}}\partial_\theta\left(\mu n\right)(\theta,t),
\end{eqnarray}
which is the fluid equation corresponding to the CTRW on the
circle. The cases $\alpha=1$ (with $\beta\equiv 0$) and $\alpha=2$ can
be obtained, respectively, as the limits $\alpha\to 1^+$ and
$\alpha\to 2^-$ of the above expressions.

\section{Propagator of the fluid limit equations in the homogeneous case}
\label{sec:propagator}

In this section we will particularize the above results to homogeneous
systems and will compute the propagator (also known as the fundamental
solution) of the fluid limit equations of the CTRW on the circle. That
is, the solution with initial condition
\begin{equation}
n(\theta,0)=\sum_{m=-\infty}^\infty\delta(\theta-2\pi m),
\end{equation}
sometimes called the {\it Dirac comb}. Its Fourier expansion is
\begin{equation}
n(\theta,0)=\frac{1}{2\pi}\sum_{m=-\infty}^\infty e^{-im\theta},
\end{equation}
i.e. $n_m(t)\vert_{t=0}=1,\forall m\in\mathbb Z$.

If the system is homogeneous Eq.~(\ref{eq:GMEperFourierinhomfl1})
takes the simple form
\begin{equation}\label{eq:GMEperFourierinhomflhomog}
\frac{\dd^{\gamma}}{\dd
t^{\gamma}} n_m(t)=A_\gamma\tau^{-\gamma}\Lambda(m)n_m(t),
\end{equation}
which after Laplace transforming allows to give the solution of the
propagator as
\begin{equation}\label{eq:mlinv}
n(\theta,t)= \frac{1}{2\pi}\sum_{m=-\infty}^\infty {\cal
L}^{-1}\left[\frac{s^{\gamma-1}}
{s^\gamma-A_\gamma\tau^{-\gamma}\Lambda(m)}\right](t) e^{-im\theta}
\end{equation}

We can make further analytical progress by using the so-called
$1$-parameter Mittag-Leffler functions, defined
by~\cite{Podlubny}:
\begin{equation}\label{eq:ml}
E_p(z) := \sum_{j=0}^\infty \frac{z^j}{\Gamma(pj+1)},
\end{equation}
for $p>0$. These functions provide a suitable generalization of the
exponential function (indeed, note that $E_1(z) = \exp(z)$) which is
useful to us because of the property:
\begin{equation}
{\cal L}^{-1}\left[\frac{s^{p-1}}{s^p \mp
a}\right](t)=E_p\left(\pm at^\gamma\right).
\end{equation}
Thus, Eq.~(\ref{eq:mlinv}) can be recast as:
\begin{equation}\label{eq:GMEperFourierinhomflhomogml}
n(\theta,t)= \frac{1}{2\pi}\sum_{m=-\infty}^\infty
E_\gamma\left(A_\gamma\Lambda(m)(t/\tau)^\gamma\right)
e^{-im\theta}.
\end{equation}

In the Markovian case, $\gamma=1$, we can produce more explicit
expressions. By setting $\gamma=1$ in
(\ref{eq:GMEperFourierinhomflhomogml}) we obtain:
\begin{equation}\label{eq:GMEperFourierinhomflhomogMark}
n(\theta,t)= \frac{1}{2\pi}\sum_{m=-\infty}^\infty
e^{\Lambda(m)t/\tau} e^{-im\theta}.
\end{equation}
In particular, if $\Lambda(m)$ is the characteristic exponent of a
stable L\'evy distribution with $\beta=0$,
\begin{equation}\label{eq:GMEperFourierinhomflhomogMarkLevy}
n(\theta,t)= \frac{1}{2\pi}\sum_{m=-\infty}^\infty
e^{(-\sigma^\alpha|m|^\alpha+i\mu m )t/\tau} e^{-im\theta}, \
\alpha\in(0,2].
\end{equation}
In the limit $t\to\infty$, $n(\theta,t)\to 1/2\pi$, as required by
the condition of conservation of probability.

Given a symmetric L\'evy distribution ($\beta=\mu=0$), it is
well-known that the fluid limit propagator in an infinite, homogeneous
system is invariant under rescaling of the so-called {\it
self-similarity variable}, $\chi:=x^\alpha/t^\gamma$, and the
propagator (with initial condition localized at the origin) at $x=0$
decays as $t^{-1/\alpha}$. Things work quite differently in a periodic
system, in which the notion of self-similiarity does not make
sense. For the time evolution of the propagator on the circle
($\beta=\mu=0$) at $\theta=0$, $n(0,t)$, there exist two distinct
regimes:

-- {\it Long-time asymptotics}: If $t>>\tau/\sigma^\alpha$ it is clear that
\begin{equation}
n(0,t)\sim \frac{1}{2\pi}\left(1+2e^{-\sigma^\alpha t/\tau}\right).
\end{equation}
Thus $n(0,t)$ decays exponentially for any $\alpha$.

-- {\it Intermediate asymptotics}: For $t<<\tau/\sigma^\alpha$, we can
   approximate the infinite sum which defines $n(0,t)$ by an integral,
\begin{equation}\label{eq:intermasymp}
n(0,t)= \frac{1}{2\pi}\sum_{m=-\infty}^\infty
e^{-\sigma^\alpha|m|^\alpha t/\tau}\approx
\frac{1}{2\pi}\int_{-\infty}^\infty e^{-\sigma^\alpha|u|^\alpha
t/\tau}\dd u \ \propto \ t^{-1/\alpha}.
\end{equation}
Hence, at short times the $\alpha$-dependent algebraic decay
characteristic of the propagator in infinite systems is recovered.

In the next subsections we work out the special cases $\alpha=2$ and
$\alpha=1$, for which we can provide analytical solutions for the
propagator.

\subsection{Diffusive transport: $\alpha=2$}

For $\alpha=2$ the step-size pdf becomes a Gaussian distribution and
(\ref{eq:GMEperFourierinhomflhomogMarkLevy}) yields:
\begin{equation}\label{eq:GMEperFourierinhomflhomogMarkLevyalpha2}
n(\theta,t)= \frac{1}{2\pi}\sum_{m=-\infty}^\infty
\exp\left(- m^2\sigma^2 t/\tau+im(\mu t/\tau-\theta)\right),
\end{equation}
which can be written in terms of the third Jacobi theta function
\cite{AbrSte}:
\begin{equation}
\vartheta_3(z,q)=1+2\sum_{m=1}^\infty q^{m^2}\cos(2mz),\quad
|q|<1,\  z\in\mathbb C.
\end{equation}
Namely,
\begin{equation}
n(\theta,t)= \frac{1}{2\pi}\vartheta_3
\left(\frac{1}{2}\left(\frac{\mu t}{\tau}-\theta\right),e^{-\sigma^2 t/\tau}\right)
\end{equation}
which is the solution of the heat equation in a periodic system.

\subsection{Cauchy distribution: $\alpha=1$}

Let us consider the case $\alpha=1$, $\beta=0$, which corresponds to a
Cauchy step-size pdf. In this situation
Eq.~(\ref{eq:GMEperFourierinhomflhomogMarkLevy}) becomes:
\begin{equation}\label{eq:gensolalpha1}
n(\theta,t)=\frac{1}{2\pi}\sum_{m=-\infty}^\infty
e^{-|m|\sigma t/\tau+im(\mu t/\tau-\theta)}.
\end{equation}
We can split the sum as
\begin{equation}\label{eq:gensolalpha12}
n(\theta,t)=\frac{1}{2\pi}\left(\sum_{m=0}^\infty e^{\left(-\sigma t/\tau
  +i(\mu t/\tau-\theta)\right)m}+ \sum_{m=1}^\infty e^{(-\sigma t/\tau -
  i(\mu t/\tau-\theta))m}\right),
\end{equation}
so that both terms are geometric series with ratios $\exp(-\sigma
t/\tau +i(\mu t/\tau-\theta))$ and $\exp(-\sigma t/\tau - i(\mu
t/\tau-\theta))$, respectively. Then,
\begin{eqnarray}
&&\sum_{m=0}^\infty e^{(-\sigma t/\tau +i(\mu t/\tau-\theta))m} =
\frac{1}{1-\exp(-\sigma t/\tau +i(\mu t/\tau-\theta))}\ ,\cr
&&\sum_{m=1}^\infty
e^{(-\sigma t/\tau - i(\mu t/\tau-\theta))m} = \frac{1}{1-\exp(-\sigma t/\tau
-i(\mu t/\tau-\theta))}-1
\end{eqnarray}
and we can give a closed expression for the propagator:
\begin{equation}\label{eq:gensolalpha13}
n(\theta,t)=
\frac{1}{2\pi}\ \frac{\sinh(\sigma t/\tau)}{\cosh(\sigma
t/\tau)-\cos(\mu t/\tau-\theta)}.
\end{equation}
It is instructive to work out the approximate form of the propagator
for small $t$ and in the vicinity of $\theta=0$ directly from
Eq.~(\ref{eq:gensolalpha13}), obtaining:
\begin{equation}
n(\theta,t)\approx \frac{1}{\pi}\ \frac{\sigma t/\tau}{(\sigma
t/\tau)^2+(\mu t/\tau-\theta)^2}.
\end{equation}
This formally coincides with the propagator in an infinite system. In
particular, $n(0,t)\propto 1/t$ for small $t$, in agreement with the
general result (\ref{eq:intermasymp}).

\section{Conclusions}\label{sec:conclusions}

Although there are numerous results concerning CTRWs in infinite
systems, much work still remains to be done on their finite-size
formulation. In this paper we have performed a detailed study of time
translationally invariant, spatially inhomogeneous, separable CTRWs on
the circle, in order to investigate how the topology can affect the
form of the Generalized Master Equation and the corresponding fluid
limit equations.

We have found that, in order to formulate the CTRW on the circle, a
non-trivial transformation of the step-size pdf must be carried out
which is strongly reminiscent of the ballooning transform familiar to
plasma physicists. During the derivation of the fluid limit of these
periodic CTRWs, it was shown that the standard Riemann-Liouville
operators need to be reformulated on the circle. We have provided
original expressions for them. In the homogeneous, Markovian case, we
have been able to compute the propagators of the resulting fluid limit
equations, even in closed form for some simple cases. Similar
calculations may be performed in the non-Markovian case, in a
straightforward manner, by using Mittag-Leffler functions. Finally, we
have also hinted how these propagators may prove useful when
determining the characteristic exponents in experimental situations,
by focusing on their short-time behaviour. We feel that the formalism
developed herein will find an application in physical systems in which
both periodicity and anomalous transport phenomena are encountered. As
mentioned in the Introduction, this could be the case of systems like
SWNT nanorings and gravitationally or magnetically confined plasmas.

\vskip 0.2cm

{\bf Acknowledgements:} Research sponsored by DGICYT (Direcci\'on
General de Investigaciones Cient\'ificas y Tecnol\'ogicas) of Spain
under Project No.~ENE2004-04319. B. A. C. gratefully acknowledges the
hospitality of CIEMAT during some of the phases of this work. Part of
this research was sponsored by the Laboratory Research and Development
Program of Oak Ridge National Laboratory, managed by UT-Battelle, LLC,
for the US Department of Energy under contract number
DE-AC05-00OR22725.


\appendix

\section{L\'evy skew alpha-stable distributions}\label{sec:appLevy}

The family of {\it L\'evy skew alpha-stable distributions} (or simply
{\it stable distributions}, or {\it L\'evy distributions}) is
parameterized by four real numbers $\alpha\in(0,2]$, $\beta\in[-1,1]$,
$\sigma>0$, and $\mu\in{\mathbb R}$. Their characteristic function
(i.e. their Fourier transform) is given by \cite{Taqqu}:
\begin{equation}\label{eq:defLevy}
\fl \hat S(\alpha,\beta,\sigma,\mu)(k)=
\left\{
\begin{array}{cc}
\exp\left(-\sigma^\alpha|k|^\alpha\left[1-i\beta\mbox{sign}(k)\tan(\frac{\pi\alpha}{2})\right]+i\mu k\right) & \alpha\neq 1,\\[8pt]
\hspace{-0.8cm}\exp\left(-\sigma|k|\left[1+i\beta\frac{2}{\pi}\mbox{sign}(k)\mbox{ln}|k|\right]+i\mu k\right) & \alpha = 1.
\end{array}
\right.
\end{equation}

According to the Generalized Central Limit Theorem
(\cite{GneKol,Taqqu}), stable distributions are the only possible
distributions with a domain of attraction. The {\it index of
stability}, $\alpha$, is related to the asymptotic behaviour of
$S(\alpha,\beta,\sigma,\mu)(x)$ at large $x$:
\begin{equation}
\fl S(\alpha,\beta,\sigma,\mu)(x)=
\left\{
\begin{array}{cc}
C_\alpha\left(\frac{1-\beta}{2}\right)\sigma^\alpha|x|^{-1-\alpha} &
x\to -\infty,\\[8pt]
C_\alpha\left(\frac{1+\beta}{2}\right)\sigma^\alpha|x|^{-1-\alpha} &
x\to \infty,
\end{array}
\right.
\end{equation}
for $\alpha\in(0,2)$. For $\alpha=2$, $S(2,\beta,\sigma,\mu)$ is a
Gaussian distribution.

The {\it skewness parameter}, $\beta$, measures the asymmetry of
the distribution. If $\beta=0$, $S(\alpha,0,\sigma,\mu)$ is
symmetric with respect to $\mu$, whereas if $\beta=\pm 1$ the
distribution is said to be {\it extremal}. Extremal distributions
when $0<\alpha<1$ are especially interesting because they are
one-sided: they are only defined for $x>0$ if $\beta=1$ and for
$x<0$ if $\beta=-1$. Their Laplace transform is given by
($\mu=0$):
\begin{equation}\label{eq:LevyextremalLaplace}
\tilde S(\alpha,1,\sigma,0)(s)=
\exp\left(-\frac{\sigma^\alpha}{\cos(\pi\alpha/2)}s^\alpha\right).
\end{equation}

\section{Fractional differential operators on the real line}
\label{sec:FDOrealine}

The {\it Riemann-Liouville fractional differential operators} are
defined as \cite{OldSpa}, \cite{Podlubny}:
\begin{eqnarray}\label{eq:RL}
&&{}_{a}D_x^\alpha f:= \frac{1}{\Gamma(m-\alpha)}\frac{\dd^m}{\dd
x^m}\int_{a}^x\frac{f(x')}{(x-x')^{\alpha-m+1}}\dd x',\cr
&&{}^{b}D_x^\alpha f:=
\frac{(-1)^{m+1}}{\Gamma(m-\alpha)}\frac{\dd^m}{\dd
x^m}\int_{x}^{b}\frac{f(x')}{(x'-x)^{\alpha-m+1}}\dd x',
\end{eqnarray}
where $\Gamma$ is the Euler Gamma function and $m\in\mathbb Z$ is
the ceiling of $\alpha$ (i.e. the smallest integer greater than or
equal to $\alpha$). When $a\to -\infty$ and $b\to\infty$ the action of the
Riemman-Liouville operators in Fourier space satisfies the important
property:
\begin{eqnarray}\label{eq:RLFourier}
&&{\cal F}[{}_{-\infty}D_x^\alpha f](k)=(-ik)^\alpha {\cal F}[f](k)\cr
&&{\cal F}[{}^{\infty}D_x^\alpha f](k)=(ik)^\alpha {\cal F}[f](k), \nonumber
\end{eqnarray}
where ${\cal F}[\cdot]$ stands for the Fourier transform. This
property justifies the name ``fractional differential operators''
because they generalize the notion of differentiation to non-integer
order.

Finally, we give the definition of the {\it Caputo fractional
differential operator} of order $\gamma\in(0,1)$ \cite{Cap}:
\begin{equation}\label{eq:Caputo}
\frac{\dd^\gamma f}{\dd t^\gamma}(t):=
\frac{1}{\Gamma(1-\gamma)}\int^t_0\frac{\dd f}{\dd t'}(t')\frac{\dd
t'}{(t-t')^\gamma},
\end{equation}
for which
\begin{equation}\label{eq:CaputoLaplace}
{\cal L}\left[\frac{\dd^\gamma f}{\dd t^\gamma}\right](s)= s^\gamma {\cal
L}[f](s)-s^{\gamma-1}f(0),
\end{equation}
where ${\cal L}[\cdot]$ denotes the Laplace transform.


\end{document}